\documentclass[a4paper,11pt,]{article}
\usepackage{pos}

\renewcommand{\vec}[1]{\boldsymbol{#1}}

\newcommand{\bfnu}{{\boldsymbol\nu}}

\newcommand{\vecP}{{\vec{P}}}
\newcommand{\Eq}{Eq.}
\newcommand{\CI}{\mathcal{I}}
\newcommand{\CK}{\mathcal{K}}
\newcommand{\CM}{\mathcal{M}}
\newcommand{\CO}{\mathcal{O}}
\newcommand{\CT}{\mathcal{T}}
\newcommand{\gothicn}{\mathfrak{n}}
\newcommand{\PartNum}{{[N]}}

\title{Analytic Expansions of Two- and Three-Particle Excited-State Energies}

\author*[a]{Dorota M.~Grabowska}
\author[b]{Maxwell T.~Hansen}
\affiliation[a]{Theoretical Physics Department, CERN, 1211 Geneva 23, Switzerland}
\affiliation[b]{Higgs Centre for Theoretical Physics, School of Physics and Astronomy, The University of Edinburgh, Edinburgh EH9 3FD, UK}
\emailAdd{dorota.grabowska@cern.ch}
\emailAdd{maxwell.hansen@ed.ac.uk}

\abstract{The last years have seen significant developments in methods relating two- and three-particle finite-volume energies to scattering observables. These relations hold for both weakly and strongly interacting systems, and studying their predictions in limiting cases can provide important cross checks as well as giving useful insights into the general formulae. In these proceedings, we present analytic results for finite-volume excited states, recovered by expanding the general relations in powers of the interaction strength. We highlight elegant patterns that emerge, especially for excited three-particle energies, and discuss various applications of the results. The two-particle results summarized here are described in more detail in Ref.~\cite{Grabowska:2021xkp}, and the three-particle results are detailed in a manuscript to appear.

  \begin{flushright}
    CERN-TH-2021-228
  \end{flushright}
}

\FullConference{%
The 38th International Symposium on Lattice Field Theory, LATTICE2021
26th-30th July, 2021
Zoom/Gather@Massachusetts Institute of Technology
}

\begin{document}
\maketitle

\section{Introduction}

Numerical lattice QCD calculations have been applied to several multi-hadron observables, including two-meson, meson-baryon and baryon-baryon scattering amplitudes as well as one-to-two decay and transition amplitudes and even three-to-three scattering amplitudes; see Refs.~\cite{Briceno:2017max, Bulava:2019hpz, Detmold:2019ghl, Edwards:2020rbo, Christ:2020kow, Aoki:2020bew, Hansen:2019nir, Rusetsky:2019gyk, Mai:2021lwb} for reviews.

The standard methodology for extracting multi-hadron observables in such calculations is to use the finite system size (the finite volume) as a probe of the infinite-volume physics. The finite spatial volume, often a periodic cubic geometry with side-length $L$, results in a discrete energy spectrum in place of the continuum of infinite-volume scattering states. These discrete energy levels retain information about the underlying scattering amplitudes, which can be extracted using field theoretic methods. A general formulation of this idea was provided in Refs.~\cite{Luscher:1986pf,Luscher:1990ux} for a system of two identical spin-zero particles with zero total momentum in the finite-volume frame. This method has since then been extended to include two-particle systems with multiple coupled channels, non-degenerate and non-identical particles, non-zero total momentum, and spin~\cite{Rummukainen:1995vs,He:2005ey,Kim:2005gf,Lage:2009zv,Bernard:2010fp,Fu:2011xz,Hansen:2012tf,Briceno:2012yi,Guo:2012hv,Briceno:2014oea}. The approach has also been extended to systems with three particles in the initial or final state \cite{Briceno:2012rv,Polejaeva:2012ut,Hansen:2014eka,Hansen:2015zga,Briceno:2017tce,Hammer:2017uqm,Hammer:2017kms,Mai:2017bge,Briceno:2018aml,Briceno:2018mlh,Jackura:2019bmu,Blanton:2019igq,Briceno:2019muc,Romero-Lopez:2019qrt,Blanton:2020gha,Blanton:2020jnm,Hansen:2020zhy,Blanton:2020gmf}.

In these proceedings, we summarize analytic relations for the two- and three-particle finite-volume energies of a system whose low-energy degrees of freedom are weakly interacting, e.g.~maximal isospin pions or kaons in QCD. The results given here focus on the low-energy regime in which only a single channel of scalar or pseudo-scalar particles can propagate and assume a $\mathbb Z_2$ symmetry that prevents coupling between even- and odd-particle-number states. The results hold for any value of total momentum $\boldsymbol P$ in the finite-volume frame but require that the corresponding non-interacting states are not accidentally degenerate. The role of accidental degeneracy for two- and three-particle states is discussed in detail in the full manuscripts \cite{Grabowska:2021xkp, Grabowska:toappear}.

We envision a number of applications for these results, including building intuition on how interactions shift finite-volume energies and exploring the convergence of contributions from higher-partial waves. Additionally, the results can be used to guide automated root finders of the full two- and three-particle quantization conditions. The results are especially useful in the three-particle sector, where implementing the full quantization condition is less straightforward.

In the following section we review the general formalism and introduce the expansion performed in this work. Then, in Secs.~\ref{sec:twopart} and \ref{sec:threepart}, we present analytic results for two- and three-particle states respectively, before briefly concluding.

\section{Formalism}

Our goal is to derive analytic expressions for finite-volume two- and three-particle energies by expanding about the non-interacting limit. We denote these energies by $E_\gothicn(L)$, where $L$ is the box length (assuming a cubic, periodic geometry) and $\gothicn$ is a collective index given by $\gothicn = \{n, \vec P, \Lambda\}$. Here the integer $n$ denotes the $n^\text{th}$ excited state in the large volume limit, $\vec P = (2 \pi/L) \boldsymbol d$ is the total spatial momentum, with $\vec d$ a three-vector of integers, and $\Lambda$ is the relevant finite-volume irreducible representation (irrep). In the limit of vanishing interactions, the energy can be expressed as the sum of finite-volume, single-particle energies.

For example, the two-particle non-interacting energy, for a channel of identical particles with mass $m$, is given by
\begin{align}
E^{(0)}_\gothicn = \omega_{\bfnu_{\mathfrak n}}+ \omega_{\vec d -\bfnu_{\mathfrak n}} \,,
\end{align}
where $\bfnu_{\mathfrak n}$ is a three-vector of integers representing the state and
\begin{equation}
\omega_{\vec n} = \sqrt{m^2 + (2 \pi/L)^2 \vec n^2} \, .
\label{eq:omegaDimLess}
\end{equation}
Similarly the non-interacting three-particle energy can be written as
\begin{align}
E^{(0)}_\gothicn = \omega_{\bfnu_{\mathfrak n, 1}}+ \omega_{\bfnu_{\mathfrak n, 2}}+ \omega_{\vec d -\bfnu_{\mathfrak n, 1}-\bfnu_{\mathfrak n, 2}} \, ,
\label{eq:ThreeNIEnergy}
\end{align}
where the state is now represented by two three-vectors, $\bfnu_{\mathfrak n, 1}$ and $\bfnu_{\mathfrak n, 2}$.

The key tool for deriving the dependence of finite-volume energies on infinite-volume scattering parameters is the quantization condition. For the present case, in which two- and three-particle states are decoupled by a symmetry, each of these sectors has its own quantization condition. Letting $N$ denote the number of particles, the general structure matches between the two cases and can be written as \cite{Luscher:1986pf, Rummukainen:1995vs, Kim:2005gf, Hansen:2014eka}
\begin{align}
\det_{\Lambda \mu} \! \Big [ \mathbb P_{\Lambda, \mu} \cdot \big ( 1 +\CI^\PartNum F^\PartNum \big ) \cdot \mathbb P_{\Lambda, \mu} \Big ] \bigg \vert_{E \, = \, E_\gothicn} = 0 \,,
\label{eq:QuantCond}
\end{align}
where $F^\PartNum$ is an $L$-dependent matrix and $\CI^\PartNum$ is an infinite-volume matrix. $F^\PartNum$ depends on both the energy and momentum in the finite-volume frame, but $\CI^\PartNum$ depends only on the Center-of-Momentum (CoM) frame energy
\begin{equation}
E^\star = \sqrt{E^2 - \boldsymbol P^2} \,.
\end{equation}
In the two-particle case, $F^{[2]}$ is a known geometric function and $\CI^{[2]}$ is the two-to-two scattering amplitude. For three particles, by contrast, $F^{[3]}$ is more complicated and depends on geometric functions as well as the two-particle scattering dynamics, and $\CI^{[3]}$ is a scheme-dependent three-particle K-matrix whose relation to the physical scattering amplitude is known.\footnote{One can also use a version of the two-particle quantization condition in which $\CI^{[2]}$ is the two-particle K-matrix and $F^{[2]}$ is modified accordingly to leave the predicted energies unchanged. See the discussion around Eq.~(98) of Ref.~\cite{Hansen:2014eka}.} The remaining quantity appearing in the quantization condition is $\mathbb P_{\Lambda, \mu}$, a projector that restricts to the irrep of interest in the relevant finite-volume group. The definition of the matrix space also depends on the value of $N$. The quantization conditions used here hold only in the range of CoM energies for which a single flavor channel can go on-shell and only up to corrections scaling as $e^{- m L}$.

Equation~\eqref{eq:QuantCond} can be used in a number of ways in practice. The simplest case arises in the two-particle sector, in a range of energies for which only a single flavor channel can contribute. Even in this case, the quantization condition involves formally infinite-dimensional matrices on the space of all possible two-particle angular momenta contributing to a given finite-volume irrep. In many practical lattice calculations, the flavor quantum numbers and the precision of lattice-determined energies are such that the systematic uncertainty of neglecting higher angular momenta is below the statistical uncertainty, so that the matrices can be truncated to a single partial wave. In such cases the two-particle scattering amplitude represents a single unknown at each value of CoM energy, and each finite-volume energy gives a determination without any need to parametrize.

The situation is more complicated whenever multiple two-particle channels can contribute and also in the three-particle sector. In such cases a parametrization of at least some of the scattering observables is often unavoidable. The optimal approach here is to identify a wide range of parametrizations and to use lattice-determined energies to identify the subset that can describe the data. The spread in successful descriptions then gives a systematic uncertainty on the extracted scattering observable. Also in this approach a truncation to a finite angular-momentum space is required.

The analytic expansions of this work are complementary to the work flows sketched above. While the expansion necessarily also requires a parametrization of the scattering observables, once this is fixed, then the truncation is automatically given by a power-counting scheme. As we describe in Ref.~\cite{Grabowska:2021xkp}, power-countings also arise for which an infinite set of angular momenta contribute at each order. We now turn to particle-number-specific details of our expansions, beginning with the two-particle sector.

\section{Two-Particle Results} \label{sec:twopart}

The explicit expressions for $\CI^{[2]}$ and $F^{[2]}$ for two identical particles, derived in Refs.~\cite{Luscher:1986pf,Luscher:1990ux,Rummukainen:1995vs, Kim:2005gf}, are
\begin{align}
\begin{split}
\CI^{[2]}_{\ell, m; \ell', m'} &= \CM_{\ell, m; \ell', m'}(E^\star) \,, \\
&= \delta_{\ell' \ell}\delta_{m' m} \frac{16 \pi E^\star}{p^\star \cot \delta_{\ell}(p^\star) - i p^\star} \,,
\label{eq:CM}
\end{split} \\
F^{[2]}_{\ell' m', \ell m}(E, \vec P, L) &= \frac 12 \lim_{\alpha \to 0^+} \bigg [\frac{1}{L^3} \sum_{\vec k} - \int_{\vec k} \bigg ] \frac{\mathcal Y_{\ell' m'}( {\vec k}^\star) \mathcal Y^*_{\ell m}( {\vec k}^\star) e^{- \alpha ( k^{\star 2} - p^{\star 2})} /(p^\star)^{\ell+\ell'}}{2 \omega_{\vec k} 2 \omega_{\vec P - \vec k}(E - \omega_{\vec k} - \omega_{\vec P - \vec k} + i \epsilon)} \,,
\label{eq:TwoPart}
\end{align}
where $E^\star$ and $p^\star$ are the energy and relative momentum in the CoM frame, related via $E^{\star 2} = 4 (p^{\star 2}+ m^2)$, $\delta_\ell$ is the scattering phase shift and $\ell, m$ are angular-momentum indices. We have also introduced $\omega_{\vec{k}} = \sqrt{\boldsymbol k^2 +m^2}$ as the on-shell time component of the four-vector $k^\mu$; we rely on context to distinguish the dimensionful and dimensionless subscripts in $\omega$. The vector $\boldsymbol k^\star$ is defined by boosting $k^\mu$ to the CoM frame, i.e.~with boost velocity $- \boldsymbol P/E$. The angular momentum is encoded via $\mathcal Y_{\ell m}(\boldsymbol x) = \sqrt{4 \pi} \vert \boldsymbol x \vert^{\ell} Y_{\ell m}(\hat {\boldsymbol x})$ where $Y_{\ell m}$ is a standard spherical harmonic.

As discussed above, to perform a general expansion of $E_\gothicn(L)$, one requires a specific parametrization of $\CM(E^\star)$, together with a power-counting scheme. The scheme assigns a power of the expansion parameter $\epsilon$ to each parameter within the amplitude, such that $\epsilon \to 0$ implies that $\CM(E^\star) \to 0$ and $E_{\mathfrak n} \to E_{\mathfrak n}^{(0)}$. The best choice of power-counting depends on the details of the system, but a useful expansion can only be performed if the low-energy scattering parameters, expressed as a positive power of length, are small in units of the box size. Again, see Ref.~\cite{Grabowska:2021xkp} for more details.

For this work, we parametrize $\CM$ using the threshold expansion
\begin{equation}
p^\star \cot \delta_{\ell}(p^\star) = - \frac{1}{a_{\ell}}\left(\frac{1}{p^\star} \right)^{\!\! 2\ell} \bigg [ 1 - \frac{r_\ell a_\ell}{2} p^{\star2} + O(p^{\star 4}) \bigg ] \,,
\label{eq:EffRangeExp}
\end{equation}
where $a_\ell, r_\ell$ are the higher partial-wave generalizations of the scattering length, $a_0$, and effective range, $r_0$. We assign a power-counting where $a_{\ell} = \mathcal O(\epsilon^{2 \ell+1})$, and assume that parameters multiplying higher powers of $ (p^{\star})^{2}$ are $\epsilon$-suppressed. With this scaling, only the $S$-wave scattering length $a_0$ enters the first few orders of the interacting energy. Thus, if we work only to these orders, the quantization condition in which $\mathcal M(E^\star)$ is set to zero for $\ell >0$ can be used.

Restricting our attention to $S$-wave only and noting that in this case, the only irrep in which shifts occur is the trivial irrep, Eq.~\eqref{eq:QuantCond} becomes
\begin{align}
p^\star \cot \delta_0(p^\star) = f(q, \vec d , L) \,,
\label{eq:SWaveQC}
\end{align}
where the function $f$ is defined by
\begin{align}
f(q, \vec d , L) \equiv - 16 \pi E^\star \, \text{Re} \big [F^{[2]}_{00,00}(E, \vec P, L) \big ] = \frac{1}{\pi L}\frac{1}{\gamma(q_\gothicn,\vec d, L)} \sum_{\vec n} \CT(\boldsymbol n \vert q_\gothicn,\vec d, L)\, \, ,
\label{eq: LittleF}
\end{align}
where $\gamma = E/E^\star$ is the Lorentz boost factor and $q$ is the dimensionless relative momentum in the CoM, defined via $(p^\star)^2 = (2\pi/L)^2 q^2$. Additionally, $\CT$ is a summand function that can be derived from the definition of $F^{[2]}$ and is presented in detail in Ref.~\cite{Grabowska:2021xkp}.\footnote{Though the symbol is not explicitly introduced, it can be inferred from Eq.~(2.26) of Ref.~\cite{Grabowska:2021xkp}.}

The expansion is conveniently performed by substituting
\begin{align}
\label{eq:qexp}
q_\mathfrak{n}(L)^2 = q_\mathfrak{n}^{(0)}(L)^2 + \sum_{j=1}^\infty \epsilon^j \, \Delta^{(j)}_{q[\mathfrak{n}]}(L) \,,
\end{align}
together with Eq.~\eqref{eq:EffRangeExp} into Eq.~\eqref{eq:SWaveQC} and solving order by order in $\epsilon$.
Once the values of $\Delta_{q[\gothicn]}^{(j)}$ are found for a given set of $j$, these can be converted to an expression for $E_{\mathfrak{n}}(L)$ via the relation
\begin{align}
q_\mathfrak{n}(L)^2 = \left(\frac{L}{2\pi}\right)^2\left[\frac{E_\mathfrak{n}(L)^2}{4} - m^2\right] \,.
\end{align}

The resulting expansion of $f(q, \vec d , L)$ is complicated by the poles arising at non-interacting finite-volume energies. To address this, we define $\vec S_{2, \gothicn} $ as the set of three-vectors entering the sum in Eq.~\eqref{eq:TwoPart} that contribute to the pole at $E_\gothicn^{(0)}$:
\begin{align}
\vec S_{2, \gothicn} = \left\{\left.\boldsymbol n \in \mathbb{Z}^3\right | E^{(0)}_\gothicn = \omega_{\boldsymbol n}+ \omega_{\vec d - \boldsymbol n} \right\} \,.
\end{align}
Breaking the infinite sum into elements of $\vec S_{2, \gothicn}$ and elements not in $\vec S_{2, \gothicn}$, we write
\begin{align}
\label{eq:fdecom}
f(q, \boldsymbol d, L) = \frac{1}{\pi L}\frac{1}{\gamma(q_\gothicn,\vec d, L)}\bigg(\,\,\underbrace{\sum_{\vec n \in \vec S_{2, \gothicn}} \CT(\boldsymbol n \vert q_\gothicn,\vec d, L)}_{\CO(1/\epsilon)}+\underbrace{\sum_{\vec n \notin \vec S_{2, \gothicn}} \CT(\boldsymbol n \vert q_\gothicn,\vec d, L)}_{\CO(1)}\,\,\bigg) \,,
\end{align}
where the indicated scaling in the underbrackets follows from substituting Eq.~\eqref{eq:qexp}. For example, the leading-order expansion in the case of $\vec P = \boldsymbol 0$ is
\begin{align}
\label{eq:fLO}
\sum_{\vec n \in \vec S_{2, \gothicn}} \CT(q_\gothicn,\vec d, L) = -\frac{g_\gothicn}{\epsilon \Delta^{(1)}_{q[\gothicn]}} + \mathcal O(\epsilon^0)\,,
\end{align}
where $g_\gothicn$ is the degeneracy factor, given by the size of $S_{2,\gothicn}$.

\begin{figure}[t]
\centering
\includegraphics[width=\textwidth]{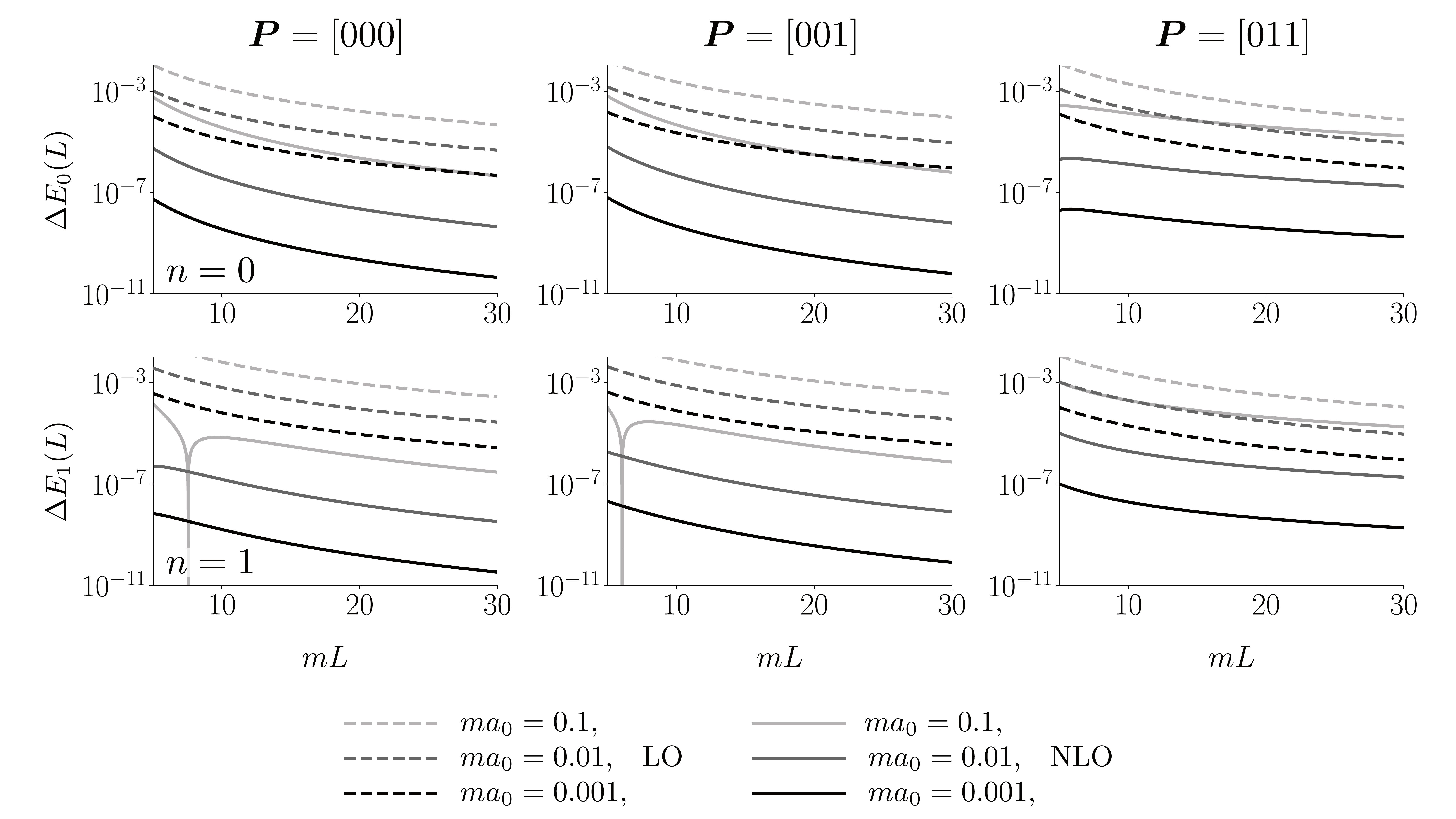}
\caption{The difference between the full finite-volume energy, which is found by numerically solving \Eq~\eqref{eq:SWaveQC}, and the analytic expression given \Eq~\eqref{eq:NLOEnergy}. We carry out the comparison in three different frames, using the shorthand that $\vec P = [d_x d_y d_z]$, and for the ground and first excited state. For each state, we subtract both the leading-order (LO) and the next-to-leading-order (NLO) energies from the full result, for three different values of the scattering length as indicated. As expected, the residual is smaller when the NLO results are used and smaller values of $m a_0$ also give smaller residuals.}
\label{fig:DeltaE}
\end{figure}

This completes the construction needed for the expansion and we now give a sample of results. For example, the resulting prediction for $E_{\mathfrak n}$ for general $\boldsymbol P$ at next-to-leading (NLO) order is
\begin{align}
E_\gothicn(L) = E^{(0)}_\gothicn (L) + g_{\mathfrak{n}} \frac{E^{(0)}_\gothicn (L)}{4 \omega_{\bfnu_\gothicn} \omega_{\vec d-\bfnu_\gothicn}}\frac{8 \pi a_0}{\gamma_\mathfrak{n}^{(0)} L^3} + \CO(\epsilon^2) \,,
\label{eq:NLOEnergy}
\end{align}
where we do not explicitly include $\epsilon$ in the NLO term. To verify this expansion, in Fig.~\ref{fig:DeltaE} we compare it to the full finite-volume energy found by numerically solving \Eq~\eqref{eq:SWaveQC}, with the parameterization of $\delta_0(p^\star)$ given in Eq.~\eqref{eq:EffRangeExp} (with $r_{0} = 0$).

Higher-order corrections can be found by expanding $f(q, \boldsymbol d, L)$ and $p^\star \cot \delta_0(p^\star)$, and thereby \Eq~\eqref{eq:SWaveQC} to higher orders in $\epsilon$ and tuning $\Delta_{q[\mathfrak n]}^{(k)}$ such that the vanishing condition is satisfied. The resulting expressions are complicated by a proliferation of terms and the appearance of geometric constants from the infinite sums within $f$. In addition, for the case of non-zero $\boldsymbol P$, the higher-order coefficients within $f(q, \boldsymbol d, L)$ are $mL$-dependent. A relatively simple expression can be given for the next-to-next-to-leading order excited state energy in the case of $\boldsymbol P = \boldsymbol 0$
\begin{align}
E_\mathfrak{n}(L)= E^{(0)}_\mathfrak{n}(L) + g_{\mathfrak{n}} \frac{E^{(0)}_\mathfrak{n} (L)}{4 \omega_{\boldsymbol{\nu}_\mathfrak{n}} \omega_{\boldsymbol{d}-\boldsymbol{\nu}_\mathfrak{n}}}\frac{8 \pi a_0}{L^3} +   g_\mathfrak{n} \frac{8 a_0^2} {E^{(0)}_\mathfrak{n}(L) L^4} \left(B_{\mathfrak{n}, 0}-\frac{4\pi^2 g_\mathfrak{n}}{E^{(0)}_\mathfrak{n}(L)^2 L^2} \right)+{\cal O}(\epsilon^3) \,,
\end{align}
where the coefficient $B_{\gothicn,0}$ is given by
\begin{align}
B_{\mathfrak{n}, 0} = \lim_{s \to -1} \sum_{\vec n \notin S_\mathfrak{n}} \Big [ q_\mathfrak{n}^{(0)2} - \vec n^2 \Big ]^{\! s} \,,
\end{align}
and the ultraviolet divergence is regulated by analytic continuation in $s$. These results match the previous work of Refs.~\cite{Huang:1957im,Luscher:1986pf,Luscher:1990ux,Beane:2007qr,Hansen:2015zta} where comparable.

The results here are only valid for finite-volume energies for which no accidental degeneracy occurs. See Ref.~\cite{Grabowska:2021xkp} for the treatment of accidentally degenerate states, which requires the inclusion of higher partial waves. We also note that the NLO energy shift can also be calculated in a power-counting scheme where no higher partial wave is suppressed and thus $\CI^{[2]}$ and $F^{[2]}$ are not truncated to finite-dimensional matrices. Again, this is described in the full manuscript.

\section{Three-Particle Results}\label{sec:threepart}

An approach similar to that summarized in the previous section can also be applied to the three-particle quantization condition. For this system, both $\CI^{[3]}$ and $F^{[3]}$ have a significantly more complicated form and also the matrix space is more complicated. In particular, $\CI^{[3]}$ is given by the divergence-free three-particle K-matrix, denoted $\mathcal K_{\text{df},3}$, and $F^{[3]}$ is given by
\begin{align}
F^{[3]} = \frac{1}{L^3}\frac{1}{2\omega}\left[\frac{F}{3} - F \frac{1}{F+G+ \CK^{-1}_2}F\right] \,,
\end{align}
where $\omega$ is a matrix of single-particle energies, $\CK_2$ is a two-particle K-matrix, with a known relation to the scattering amplitude, and $F$ and $G$ are finite-volume functions. These expressions were first introduced in Ref.~\cite{Hansen:2014eka}. Rather than being matrices only on the two-particle angular momentum space $\ell' m', \ell m$, as in the previous section, these matrices now act on the Kronecker product of this space together with a finite-volume spectator momentum $\boldsymbol k = (2 \pi /L) \boldsymbol n$, abbreviated $k$. The full index space is thus denoted $k', \ell', m';\, k, \ell, m$.

For example, $G$ is given by
\begin{align}
G_{k', \ell', m';\, k, \ell, m} \equiv\frac{1}{2\omega_{\boldsymbol k} L^3} \left(\frac{1}{q^\star_{k'}}\right)^{\ell'}\frac{\mathcal Y_{\ell', m'}(\boldsymbol k^\star) \mathcal Y^*_{\ell, m}(\boldsymbol {k}'^\star) H(\vec k, \vec k')}{2 \omega_{\vecP - \vec k- \vec k'}\left(E- \omega_{\vec k'} - \omega_{\vec k} - \omega_{\vecP -\vec k- \vec k'}+ i \epsilon\right)} \left(\frac{1}{q^\star_{k}}\right)^\ell \,,
\label{eq:GDef}
\end{align}
where $\boldsymbol k^\star$ is defined by boosting $k^\mu = (\omega_{\boldsymbol k}, \boldsymbol k)$ with boost velocity $\beta_{k'} \equiv - (\vec P- \vec k')/(E- \omega_{\vec k'})$, and $\boldsymbol k'^\star$ is defined with primed and unprimed objects exchanged everywhere. Here $q^\star_k$ is the relative momentum in the CoM frame for the two-particle system where the particle with momentum $\vec k$ is the spectator:
\begin{align}
(q^\star_k)^2 \equiv E^{\star 2}_{2, k}/4 - m^2 \,, \qquad E^{\star 2}_{2, k} = \left(E - \omega_{\boldsymbol k}\right)^2 -\left(\vec P - \vec k\right)^2 \, .
\end{align}
In slight notational tension to the two-particle case, this is a dimensionful quantity, following the convention of Ref.~\cite{Hansen:2014eka}. Lastly, $H(\vec k, \vec k')$ is a smooth cut-off function that vanishes for $ E^{\star 2}_{2, k}  < 0$ or  $ E^{\star 2}_{2, k'}  < 0$, and is equal to unity for $ E^{\star 2}_{2, k} > 4 m^2 $ and $ E^{\star 2}_{2, k'} > 4 m^2 $. More details are given, for example, in Refs.~\cite{Hansen:2014eka,Hansen:2016fzj}.

\begin{figure}[t]
\centering
\includegraphics[width=0.95\textwidth]{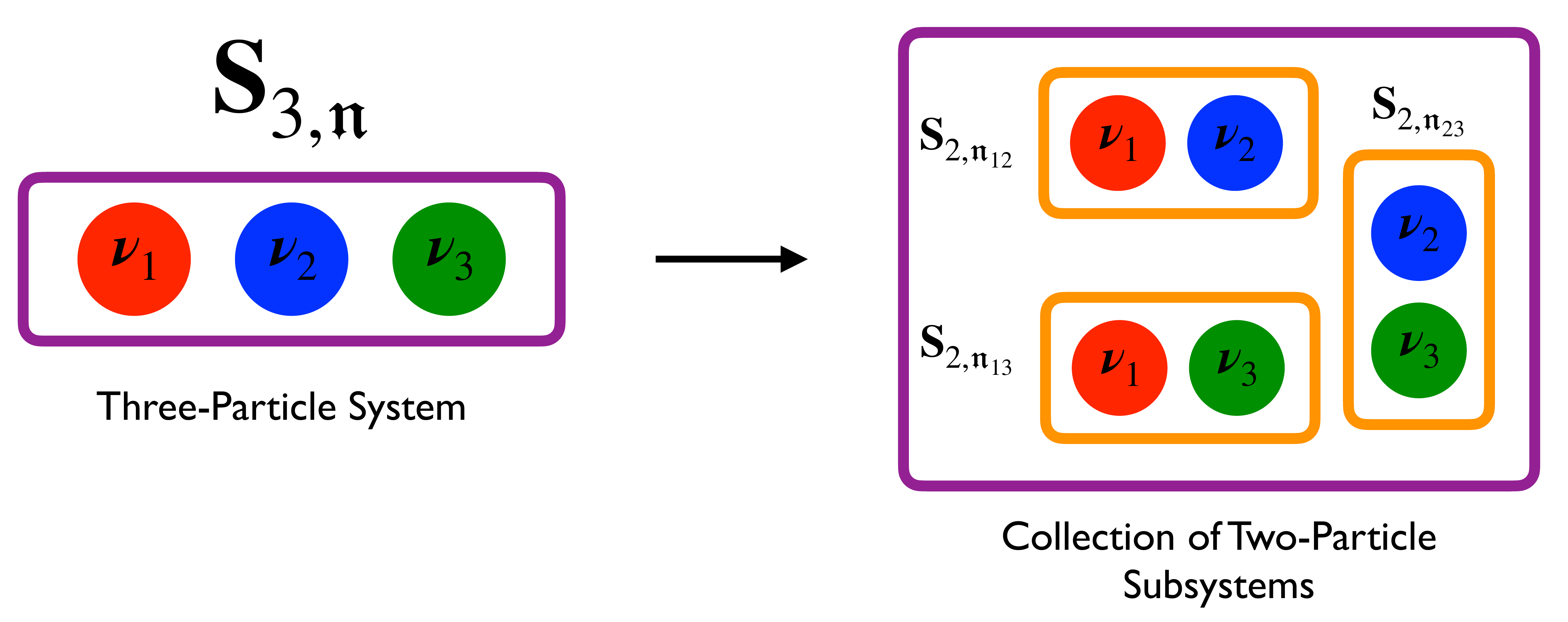}
\caption{Decomposition of a three-particle system (with no accidental degenreacy) into a collection of three two-particle subsystems. For each set $S_{2, \gothicn_i}$, the omitted particle is called the spectator. This partitioning forms the backbone of our expansion of the three-particle quantization condition.}
\label{fig:NonDegSch}
\end{figure}

To find the NLO energy, we make use of three key facts, all of which are expanded upon in the forthcoming manuscript \cite{Grabowska:toappear}. The first is that the NLO energy is found by solving for the pole in $F^{[3]}$, which occurs when
\begin{align}
\det \left[F+G+ \CK^{-1}_2 \right]= 0 \,.
\label{eq:NLOQC}
\end{align}
This statement was already demonstrated in Ref.~\cite{Hansen:2016fzj} for the ground state in the case of $\vec P = \boldsymbol 0$. Note that this immediately implies that the NLO three-particle energy depends only on the two-particle scattering dynamics. 

Second we note that, as in the two-particle case, the matrices $F$ and $G$ have poles at the non-interacting three-particle energies. This can be seen explicitly in the definition of $G$ given in \Eq~\eqref{eq:GDef}. This motivates the introduction of the three-particle analog of $S_{2,\mathfrak{n}}$:
\begin{align}
S_{3,\mathfrak{n}} = \left\{\vec n_1 \in \mathbb{Z}^3 \bigg| E^{(0)}_{3, \mathfrak{n}} -\left(\omega_{\vec n_1}+\omega_{\vec n_2}+\omega_{\vec d - \vec n_1-\vec n_2}\right) = 0 \quad \forall \,\, \vec n_2 \in \mathbb{Z}^3\right\} \,,
\end{align}
where $E^{(0)}_{3, \mathfrak{n}}$ is the three-particle, non-interacting energy of interest, see \Eq~\eqref{eq:ThreeNIEnergy}. Note that, as long as the three-particle state is non-degenerate as we assume here, then $\vec S_{3, \gothicn}$ is equal to the union of at most three distinct two-particle sets, $\vec S_{2, \gothicn_i}$. This is shown schematically in Fig.~\ref{fig:NonDegSch}.

The third key fact concerns the textures of the leading order expansions of the matrices $\CK_2$, $F$ and $G$. This relies on expanding the matrices following the same power-counting scheme as above with the two-particle scattering length $a_0 = \mathcal O(\epsilon)$ and with $E_{3, \mathfrak{n}}  - E^{(0)}_{3, \mathfrak{n}}  = \mathcal O(\epsilon)$.\footnote{In the three-particle case we take the expansion directly in terms of $E_{3, \mathfrak{n}}$, since a unqiue relative momentum cannot be defined with three particles.} For this power-counting, only the $S$-wave contributes at NLO and the matrices can be truncated to the space of spectator momenta only: $k', k$. Keeping the $\mathcal O(1/\epsilon)$ contributions from $\CK^{-1}_2$, $F$ and $G$ one can classify all possible matrix textures, on this space, based on the particular three-particle state of interest. In particular, $\CK^{-1}_2$ and $F$ are always diagonal and $G$ has one of five textures, summarized by
\begin{align}
G \sim \left\{\left(\begin{array}{c|c|c}
\lozenge[n_{a}]&\blacklozenge[g_{\gothicn a}/2]&\blacklozenge[g_{\gothicn a}/2]\\ \hline \blacklozenge[g_{\gothicn b}/2] &\lozenge[n_{b}] &\blacklozenge[g_{\gothicn b}/2]\\ \hline
\blacklozenge[g_{\gothicn c}/2]&\blacklozenge[g_{\gothicn c}/2]&\lozenge[n_{c}]
\end{array}\right),\,\, \left(\begin{array}{c|c}
\lozenge[n_{a}]&\blacklozenge[g_{\gothicn a}]\\ \hline
\blacklozenge[g_{\gothicn b}/2] &\blacklozenge[g_{\gothicn b}/2]
\end{array}\right),\,\, \left(\begin{array}{c|c}
\lozenge[n_{a}]&\blacklozenge[1]\\ \hline
\blacklozenge[1] &\blacklozenge[1] \end{array}\right),\,\,\blacklozenge[ g_{\gothicn a}], \,\,\blacklozenge[1] \right\} \,,
\end{align} 
where the symbols are defined as
\begin{align}
\lozenge[n]&: \quad \text{square matrix of all zeros, with dimensions } n 
\times n \nonumber \,, \\
\blacklozenge[n]&: \quad \text{rectangular matrix of zeros and ones, with } n \text{ non-zero entries per row} \,,
\end{align}
and $g_{\gothicn i}$ is the size of set $\vec S_{2,\gothicn_i}$. Additionally, $n_i$ is the number of vectors in $S_{3, \gothicn}$ that are related to each other by elements of the point group defined by $\vec P$. These vectors play the role of the spectator particle when defining the set $S_{2, \gothicn_i}$.

Combining the three key facts we have summarized, it is possible to show that the quantization condition of Ref.~\cite{Hansen:2014eka} predicts that the NLO energy for a three-particle state is given by
\begin{align}
\label{eq:finalresult}
E_{3, \mathfrak{n}}=E^{(0)}_{3, \mathfrak{n}} + \sum_{i = 1}^3 \Delta^{(1)}_{E[2,\mathfrak{n}_i]}+ {\cal O}(\epsilon^2)  \,, \qquad \Delta^{(1)}_{E[2,\mathfrak{n}]}\equiv g_{\mathfrak{n}} \frac{E^{(0)}_\mathfrak{n} (L)}{4 \omega_{\boldsymbol{\nu}_\mathfrak{n}} \omega_{\boldsymbol{d}-\boldsymbol{\nu}_\mathfrak{n}}}\frac{8 \pi a_0}{\gamma_\mathfrak{n}^{(0)} L^3} \,,
\end{align}
where $ \Delta^{(1)}_{E[2,\mathfrak{n}]}$ is recognized as the leading-order energy shift in the two-particle system, given in \Eq~\eqref{eq:NLOEnergy}.

The interpretation of this result is straightforward: \textit{\textbf{the leading-order energy shift of any non-degenerate three-particle system is the sum of leading-order energy shifts in all the two-particle subsystems}}. This highly intuitive result has been derived for the ground state in the $\vec P = \boldsymbol 0$ case in Refs.~\cite{Huang:1957im,Beane:2007qr,Hansen:2015zta,Hansen:2016fzj,Muller:2020vtt}, but the result for all excited states and all moving frames is new to this work.
A comparison of this generic expression to specific excited state results derived using a different formalism \cite{Pang:2019dfe,Romero-Lopez:2020rdq}, together with the full proof, is given in a manuscript in preparation \cite{Grabowska:toappear}.

\section{Conclusions}

In these proceedings, we have summarized analytic relations for two- and three-particle finite-volume energies of weakly interacting systems, focusing on the low-energy regime in which only a single channel of scalar or pseudo-scalar particles can propagate. The relations are valid for any excited state, provided the latter does not have an accidental degeneracy, and for any value of the total momentum. The work summarized here is described in more detail in Refs.~\cite{Grabowska:2021xkp, Grabowska:toappear}.

The most significant result presented in these proceedings is that the leading shift in a generic, excited state three-particle energy is given by summing the shifts associated to all two-particle pairs, as summarized in Eq.~\eqref{eq:finalresult}. This holds provided that no accidental degeneracies are present in the state of interest. Although the result is very intuitive, deriving it from the full three-particle quantization condition is highly non-trivial. We also view this as another cross-check on the formalism itself, though to make this solid an independent derivation for a generic three-particle excited state is required. This could be done, for example, by explicitly considering the excited states in $\lambda \phi^4$ theory, as was done in Ref.~\cite{Hansen:2015zta} for the ground state. Additional future work, going beyond the manuscript in preparation, includes treating multiple channels and channels with non-identical and non-degenerate particles.

\acknowledgments
We thank Fernando Romero-L{\'o}pez and Steve Sharpe for useful discussions. D.M.G. would additionally like to thank Michael Wagman for useful discussions. The work of M.T.H.~is supported by UK Research and Innovation Future Leader Fellowship MR/T019956/1.

\bibliographystyle{JHEP}
\bibliography{refs.bib}

\end{document}